\documentclass[11pt]{article}

\usepackage[utf8]{inputenc}
\usepackage[a4paper,margin=2.5cm]{geometry}
\usepackage{amsmath,amssymb,amsthm,bm}
\usepackage{graphicx}
\usepackage{booktabs}
\usepackage[version=4]{mhchem}
\usepackage{siunitx}
\usepackage{xcolor}
\usepackage{authblk}
\usepackage{microtype}
\usepackage[numbers,sort&compress]{natbib}
\usepackage[colorlinks=true,linkcolor=blue!50!black,citecolor=blue!50!black,urlcolor=blue!50!black]{hyperref}
\usepackage[capitalise,nameinlink]{cleveref}

\crefname{equation}{Eq.}{Eqs.}
\crefname{figure}{Fig.}{Figs.}
\crefname{section}{Sec.}{Secs.}
\crefname{table}{Table}{Tables}
\crefname{proposition}{Proposition}{Propositions}

\theoremstyle{plain}
\newtheorem{proposition}{Proposition}

\newcommand{\vect}[1]{\mathbf{#1}}
\newcommand{\T}{\mathsf{T}}
\newcommand{\dt}{\Delta t}
\newcommand{\code}[1]{\texttt{#1}}

\title{Stable ring-polymer molecular dynamics without Fourier transforms:\\
a linear-scaling bead-space Cayley propagator}

\author[1]{Yuhao Chen}
\author[2]{Yongle Li}
\author[1,3,4]{Xiao He\thanks{Corresponding author. E-mail: \href{mailto:xiaohe@phy.ecnu.edu.cn}{xiaohe@phy.ecnu.edu.cn}}}
\affil[1]{Shanghai Engineering Research Center of Molecular Therapeutics and New Drug
  Development, Shanghai Frontiers Science Center of Molecule Intelligent Syntheses,
  School of Chemistry and Molecular Engineering, East China Normal University,
  Shanghai 200062, China}
\affil[2]{Department of Physics, International Center of Quantum and Molecular
  Structures, and Shanghai Frontiers Science Center of Quantum and Superconducting
  Matter States, Shanghai University, Shanghai 200444, China}
\affil[3]{Anhui Provincial Key Laboratory of Advanced Catalysis and Energy Materials,
  Anhui Ultra High Molecular Weight Polyethylene Fiber Engineering Research Center,
  School of Chemistry and Chemical Engineering, Anqing Normal University,
  Anqing 261433, China}
\affil[4]{New York University--East China Normal University Center for Computational
  Chemistry, New York University Shanghai, Shanghai 200062, China}

\date{}

\begin{document}
\maketitle

\begin{abstract}
Ring-polymer molecular dynamics (RPMD) builds zero-point energy and tunnelling into
classical trajectories and is widely used for bimolecular rate coefficients. At low
temperature it requires many beads, and the stiff ring-polymer springs then limit the
integrator. The Cayley modification of the free ring-polymer step removes this limit,
but it has so far been formulated mode by mode, which requires Fourier transforms of
every degree of freedom at every step. Here we show that, in bead coordinates, the
Cayley step reduces to one cyclic tridiagonal solve for the midpoint of the step, with
a matrix that depends only on the number of beads $P$, the temperature and the time
step, and not on the atomic masses. The bead-space step is exactly the normal-mode
Cayley map: it conserves the free ring-polymer energy, preserves the free ring-polymer
Boltzmann distribution, and is stable for any $P$ whenever $\Omega\Delta t<2$ for the
highest physical frequency $\Omega$. A Thomas recursion with a rank-two Woodbury
correction solves it in about $18P$ operations per degree of freedom without pivoting,
with a condition number bounded by $1+(\omega_P\Delta t)^2$, and the inverse matrix
decays geometrically with bead separation, so the step is local in bead index.
Implemented in the open-source PyRPMD package, the propagator reproduces normal-mode
Cayley to $1.5\times10^{-15}$ for up to 2048 beads and its cost grows linearly with $P$.
For direct RPMD (dRPMD), which has so far integrated the springs explicitly, these
properties raise the stability limit on the time step 23--29-fold in published
low-temperature settings; at \SI{20}{K} with 3072 beads, the Cayley step conserves the
ring-polymer energy as well as velocity Verlet does with a time step about ten times larger,
while leaving the low-lying internal modes that govern spurious resonances essentially
unchanged.
\end{abstract}

\section{Introduction}
\label{sec:intro}

Many chemical reactions show pronounced nuclear quantum effects at and below room
temperature: zero-point energy shifts effective barriers, and tunnelling of light
atoms drives rates away from classical predictions. Ring-polymer molecular dynamics
(RPMD) captures these effects within a classical-trajectory framework by replacing
each atom with a ring of $P$ beads, and has become a standard tool for gas-phase
bimolecular rate coefficients.\cite{craig2004,craig2005a,craig2005b,habershon2013,suleimanov2016}

The difficulty grows as the temperature falls. Converging quantum effects at low
temperature requires more beads, and the frequency of the springs that connect them
grows linearly with $P$. The standard propagator transforms the springs into normal
modes and rotates each mode exactly,\cite{ceriotti2010pile} but combined with the
force impulses this splitting becomes resonantly unstable once the highest internal
mode turns through about half a period in one time step.\cite{korol2019}

Korol \textit{et al.}\ removed this instability by replacing the exact rotation with
a Cayley transform, which is stable for any internal-mode frequency while remaining
symplectic and preserving the free ring-polymer equilibrium
distribution.\cite{korol2019,korol2020,rosaraices2021} Gui \textit{et al.}\
implemented the Cayley step in RPMDrate and showed for \ce{OH + CH4} that the
time step could be raised from \SI{0.1}{fs} to \SI{0.5}{fs} at high temperature with
rate coefficients within about 10\%.\cite{gui2022} These formulations remain in the
normal-mode framework and therefore transform every degree of freedom to and from
normal modes at every step. Korol \textit{et al.}\ noted that the Cayley step can also
be written in bead coordinates;\cite{korol2019} the structure of the resulting linear
system, and what that structure implies for cost, conditioning, locality and
parallelism, have not been worked out. The question matters most for direct RPMD
(dRPMD), which simulates reactant collisions directly and is used for barrierless and
complex-forming reactions at low temperature, such as \ce{H2 + H3+}, \ce{OH + CH3OH}
and \ce{OH + H2CO}.\cite{suleimanov2018,delmazo2019,delmazo2021} dRPMD runs with
thousands of beads distributed over processes and integrates the springs explicitly,
which confines the time step to about \SI{0.1}{fs} at every temperature.\cite{delmazo2021}

Here we formulate the Cayley step in bead coordinates and derive its properties. The
step is a single nearest-neighbour linear system for the midpoint of the step: each
bead couples to its two neighbours, so the coefficient matrix is tridiagonal apart
from two corner elements that close the ring, and it is independent of atomic mass.
We show that this bead-space step is the normal-mode Cayley map itself, so it inherits
every stability and conservation property of that map; that it is solved in
$\mathcal O(P)$ operations by a pivot-free recursion with bounded conditioning; and
that its inverse is exponentially local in bead index. Applied to dRPMD, these results
raise the stability limit on the time step by more than an order of magnitude, leave
the physics of spurious resonances intact, and preserve nearest-neighbour
communication. The propagator is the default integrator of the PyRPMD
package,\cite{pyrpmd} which we use to verify each property numerically.
\Cref{sec:background} reviews RPMD rate theory and the Cayley step,
\cref{sec:beadspace} develops the bead-space formulation, \cref{sec:drpmd} derives its
consequences for dRPMD, \cref{sec:numerics} verifies the central results, and
\cref{sec:conclusions} summarizes the implications.

\section{RPMD and the Cayley free ring-polymer step}
\label{sec:background}

\subsection{RPMD rate theory}
\label{sec:rpmd}

RPMD rate theory and its implementation in RPMDrate have been described in
detail;\cite{collepardo2009,suleimanov2013} we summarize the elements used here. For
$N$ atoms, each represented by $P$ beads, the RPMD Hamiltonian is
\begin{equation}
  H_P=\sum_{j=1}^{N}\sum_{k=1}^{P}\left[\frac{p_{j,k}^{2}}{2m_j}
  +\frac{1}{2}m_j\,\omega_P^{2}\left(q_{j,k}-q_{j,k+1}\right)^{2}\right]
  +\sum_{k=1}^{P}V\!\left(q_{1,k},\dots,q_{N,k}\right),\qquad q_{j,P+1}\equiv q_{j,1},
  \label{eq:HP}
\end{equation}
with $\omega_P=P/(\beta\hbar)$ and $\beta=1/(k_\mathrm{B}T)$. The rate coefficient is
evaluated in the Bennett--Chandler form as the product of a transmission coefficient
and a centroid-density quantum transition-state-theory rate,\cite{collepardo2009,suleimanov2013}
\begin{equation}
  k(T)=\kappa(t\to t_p)\,k_{\mathrm{QTST}}(T;\xi^{\ddagger}).
  \label{eq:BC}
\end{equation}
Here $k_\mathrm{QTST}$ follows from the centroid potential of mean force along the
reaction coordinate $\xi$, obtained by umbrella sampling, and $\kappa$ is computed
from recrossing trajectories launched at the dividing surface $\xi^\ddagger$. Both
parts integrate the same Hamiltonian: the first with thermostatted dynamics, the
second with microcanonical dynamics.

The dynamics is split into the free ring polymer $H_0$ (kinetic energy plus springs)
and the physical potential. One step follows the symmetric sequence B--A--B: a
half-step impulse from $-\nabla V$ (B), a full free ring-polymer step (A), and a
second half-step impulse. In umbrella sampling RPMDrate adds an
Andersen\cite{andersen1980} or GLE\cite{ceriotti2010gle} thermostat, and the parent
trajectory constrains the centroid to the dividing surface with SHAKE/RATTLE; the
recrossing stage is unthermostatted.\cite{suleimanov2013} $H_0$ separates into
independent one-dimensional problems, one for each Cartesian component of each atom.
With bead positions $\vect q$ and velocities $\vect v=\vect p/m$,
\begin{equation}
  \dot{\vect q}=\vect v,\qquad \dot{\vect v}=-\omega_P^{2}\,\vect L\,\vect q,\qquad
  \vect L=\begin{pmatrix}
    2&-1&0&\cdots&-1\\
    -1&2&-1&\cdots&0\\
    \vdots&\ddots&\ddots&\ddots&\vdots\\
    0&\cdots&-1&2&-1\\
    -1&\cdots&0&-1&2
  \end{pmatrix}.
  \label{eq:free}
\end{equation}
The mass has dropped out of \cref{eq:free}; this is the starting point of the
bead-space formulation in \cref{sec:beadspace}.

\subsection{Normal-mode propagation and the Cayley step}
\label{sec:cayley}

The conventional RPMD propagator inherits the normal-mode scheme of path-integral
molecular dynamics.\cite{ceriotti2010pile} $\vect L$ is circulant and is
diagonalized by the real discrete Fourier transform, $\vect L=\vect C\bm\Lambda\vect C^\T$,
with eigenvalues $\lambda_k=4\sin^2(\pi k/P)$ and internal-mode frequencies
$\omega_k=2\omega_P\sin(\pi k/P)$. For mode $k$, let $x_k=\omega_k\dt$ and
$\vect A_k=\begin{pmatrix}0&1\\-\omega_k^2&0\end{pmatrix}$; the exact normal-mode step
is the rotation $\exp(\dt\,\vect A_k)$. Combined with the force impulses, this
rotation is not strongly stable. For a harmonic external force of frequency $\Omega$,
the half-trace of the one-step B--A--B map of mode $k$ is
$\cos x_k-(\Omega\dt)^2\sin(x_k)/(2x_k)$, which exceeds unity in magnitude in a
window just below every $x_k=n\pi$.\cite{korol2019} Since the largest $\omega_k$ is
$2\omega_P$, at \SI{300}{K} and $\dt=\SI{0.5}{fs}$ this condition is reached already
at $P\approx80$.

Korol \textit{et al.}\ replaced the exact rotation with the Cayley transform,\cite{korol2019}
\begin{equation}
  \operatorname{cay}(\dt\,\vect A_k)
  =\left(\vect I-\tfrac{\dt}{2}\vect A_k\right)^{-1}\left(\vect I+\tfrac{\dt}{2}\vect A_k\right)
  =\frac{1}{1+x_k^{2}/4}
  \begin{pmatrix}1-x_k^{2}/4&\dt\\-\omega_k^{2}\dt&1-x_k^{2}/4\end{pmatrix}.
  \label{eq:cayley}
\end{equation}
This is a rotation by $\theta_k=2\arctan(x_k/2)<\pi$, which never reaches resonance
for any internal mode. The Cayley map is symplectic and time-reversible and leaves the
free ring-polymer Boltzmann distribution invariant.\cite{korol2019,korol2020,rosaraices2021}
With the force impulses included, the half-trace becomes
\begin{equation}
  \frac{1-x_k^2/4-\Omega^2\dt^2/2}{1+x_k^2/4},
  \label{eq:halftrace}
\end{equation}
and the step is stable for all modes and all $P$ if and only if $\Omega\dt<2$, the
stability limit of classical velocity Verlet.

Rate calculations use the Cayley step in different compositions. Gui \textit{et al.}\
used a BCOCB scheme in umbrella sampling and B--C--B in recrossing trajectories, where
C denotes the Cayley free ring-polymer step and O the thermostat.\cite{gui2022} Korol
\textit{et al.}\ showed that the two half-steps of BCOCB require the square-root form
of the Cayley map to remain strongly stable.\cite{korol2020} Here C always denotes the
full Cayley step: deterministic dynamics is B--C--B, and thermostatted sampling adds an
Andersen thermostat between the impulses.

For a linear system, the Cayley map is the implicit midpoint rule,\cite{hairer2006}
\begin{equation}
  \vect q'=\vect q+\frac{\dt}{2}\left(\vect v+\vect v'\right),\qquad
  \vect v'=\vect v-\dt\,\omega_P^{2}\,\vect L\,\bar{\vect q},\qquad
  \bar{\vect q}=\tfrac12\left(\vect q+\vect q'\right),
  \label{eq:midpoint}
\end{equation}
that is, the spring force is evaluated at the midpoint $\bar{\vect q}$ of the step.
Existing formulations apply \cref{eq:cayley} mode by mode; the next section treats
\cref{eq:midpoint} directly in bead coordinates.

\section{Bead-space formulation of the Cayley step}
\label{sec:beadspace}

\subsection{The Cayley step as a cyclic tridiagonal system}
\label{sec:system}

Inserting $\vect v'$ from \cref{eq:midpoint} into the definition of $\bar{\vect q}$
gives $\bar{\vect q}=\vect q+\tfrac{\dt}{2}\vect v-\alpha\vect L\bar{\vect q}$, so the
midpoint satisfies the linear system
\begin{equation}
  \vect M\,\bar{\vect q}=\vect q+\frac{\dt}{2}\,\vect v,\qquad
  \vect M=\vect I+\alpha\vect L,\qquad
  \alpha=\frac{(\omega_P\dt)^{2}}{4},
  \label{eq:system}
\end{equation}
after which the new positions and momenta follow explicitly,
\begin{equation}
  \vect q'=2\bar{\vect q}-\vect q,\qquad
  \vect p'=\vect p-\dt\,m\,\omega_P^{2}\,\vect L\,\bar{\vect q}.
  \label{eq:update}
\end{equation}
The right-hand side of \cref{eq:system} involves no neighbour coupling, and the only
coupling in \cref{eq:update} is the nearest-neighbour stencil of $\vect L$.

\begin{figure}[t]
  \centering
  \includegraphics[width=\textwidth]{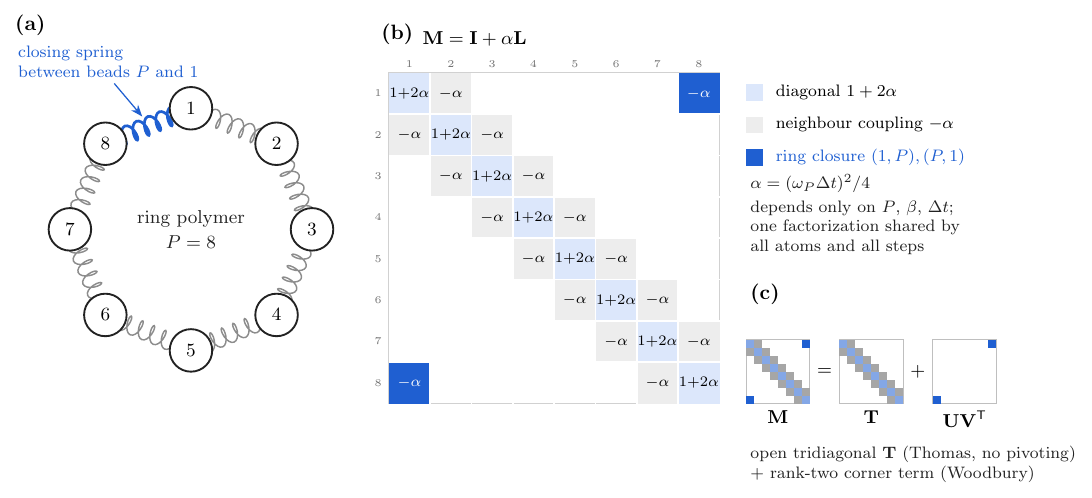}
  \caption{Structure of the Cayley step in bead coordinates.
  (a)~Ring polymer with $P=8$ beads; the closing spring between beads $P$ and $1$ is
  highlighted.
  (b)~Coefficient matrix $\vect M=\vect I+\alpha\vect L$ of \cref{eq:system}:
  neighbouring beads give the tridiagonal band, and the closing spring gives the two
  corner elements.
  (c)~Splitting used by the solver, $\vect M=\vect T+\vect U\vect V^\T$, into the open
  tridiagonal part $\vect T$ and a rank-two corner term (\cref{eq:woodbury}).}
  \label{fig:schematic}
\end{figure}

The structure of $\vect M$ mirrors the topology of the ring (\cref{fig:schematic}).
Each bead couples only to its two neighbours, so $\vect M$ is tridiagonal; closing the
ring adds just the two corner elements. $\vect M$ contains only the number of beads,
the temperature and the time step. It is independent of atomic mass, because writing
the equations in velocities rather than momenta removes the mass. One matrix therefore
serves every Cartesian component of every atom, and it changes only when $P$, $\beta$
or $\dt$ does.

\subsection{Equivalence with normal-mode Cayley}
\label{sec:equivalence}

\begin{proposition}\label{prop:equivalence}
The map $(\vect q,\vect p)\mapsto(\vect q',\vect p')$ defined by
\cref{eq:system,eq:update} coincides with the normal-mode Cayley step
\cref{eq:cayley} applied to every mode. Consequently it conserves $H_0$ exactly,
propagates the centroid as exact free flight, preserves the free ring-polymer Boltzmann
distribution, and, composed with force impulses, is stable for all $P$ if and only if
$\Omega\dt<2$.
\end{proposition}

\begin{proof}
$\vect M$ and $\vect L$ are diagonalized by the same Fourier transform. Multiplying
\cref{eq:system} by $\vect C^\T$ decouples it mode by mode,
$\tilde{\bar q}_k=(\tilde q_k+\tfrac{\dt}{2}\tilde v_k)/(1+x_k^2/4)$, where we used
$\alpha\lambda_k=x_k^2/4$. Then
\begin{equation*}
  \tilde q_k'=2\tilde{\bar q}_k-\tilde q_k=\frac{(1-x_k^2/4)\,\tilde q_k+\dt\,\tilde v_k}{1+x_k^2/4},\qquad
  \tilde v_k'=\tilde v_k-\dt\,\omega_k^2\,\tilde{\bar q}_k
  =\frac{(1-x_k^2/4)\,\tilde v_k-\omega_k^2\dt\,\tilde q_k}{1+x_k^2/4},
\end{equation*}
which are the two rows of \cref{eq:cayley}. The implicit midpoint rule conserves every quadratic invariant of a
linear system,\cite{hairer2006} which gives exact conservation of $H_0$; for the
centroid ($\lambda_0=0$) the update reduces to $\tilde q_0'=\tilde q_0+\dt\,\tilde v_0$.
Invariance of the Boltzmann distribution and the stability criterion follow from the
corresponding properties of \cref{eq:cayley} and \cref{eq:halftrace}.\cite{korol2019}
\end{proof}

The bead-space step is therefore not an approximation to normal-mode Cayley but a
different algorithm for the same map; the choice between them is a matter of cost and
structure, which the remainder of this section addresses.

\subsection{An \texorpdfstring{$\mathcal O(P)$}{O(P)} pivot-free solver}
\label{sec:algorithm}

The matrix in \cref{eq:system} depends only on $P$, $\beta$ and $\dt$, so it can be
factorized once and shared. We split $\vect M$ into its open tridiagonal part $\vect T$
and a rank-two corner term and apply the Woodbury identity,\cite{hager1989}
\begin{equation}
\begin{gathered}
  \vect M=\vect T+\vect U\vect V^{\T},\qquad
  \vect U=-\alpha\left[\vect e_1,\ \vect e_P\right],\qquad
  \vect V=\left[\vect e_P,\ \vect e_1\right],\\
  \vect M^{-1}\vect r=\vect y-\vect Z\,\vect S^{-1}\vect V^{\T}\vect y,\qquad
  \vect y=\vect T^{-1}\vect r,\qquad
  \vect Z=\vect T^{-1}\vect U,\qquad
  \vect S=\vect I_2+\vect V^{\T}\vect Z.
\end{gathered}
\label{eq:woodbury}
\end{equation}
Unlike the classical rank-one cyclic Thomas variant,\cite{temperton1975} this splitting
leaves the diagonal of $\vect T$ unchanged, so $\vect T$ keeps the diagonal dominance of
$\vect M$. One step proceeds as follows.
\begin{enumerate}
  \item \emph{Setup (once per $P$, $\beta$, $\dt$).} Thomas recursion $d_1=1+2\alpha$,
    $\ell_k=-\alpha/d_{k-1}$, $d_k=1+2\alpha+\alpha\ell_k$; store $1/d_k$, $\ell_k$,
    the two columns $\vect z_1,\vect z_2$ of $\vect Z$, and the four elements of
    $\vect S^{-1}$: $4P+4$ numbers shared by all atoms.
  \item \emph{Right-hand side.} $r_k=q_k+\tfrac12\dt\,v_k$.
  \item \emph{Tridiagonal solve.} $y_k\leftarrow r_k-\ell_k y_{k-1}$, then
    $y_k\leftarrow(y_k+\alpha y_{k+1})/d_k$.
  \item \emph{Corner correction.} $(c_1,c_2)^\T=\vect S^{-1}(y_P,y_1)^\T$,
    $\bar q_k=y_k-c_1z_{1,k}-c_2z_{2,k}$.
  \item \emph{Update.} $q_k'=2\bar q_k-q_k$ and
    $p_k'=p_k-\dt\,m\,\omega_P^2(2\bar q_k-\bar q_{k-1}-\bar q_{k+1})$, with periodic
    indices.
\end{enumerate}
Steps 2--5 cost about $18P$ floating-point operations per degree of freedom. A
normal-mode step needs four real FFTs plus the mode rotations, about $10P\log_2P+6P$
operations, and a dense transform about $8P^2+6P$ (\cref{sec:cost}). Because every
degree of freedom uses the same coefficients, the $3N$ Cartesian rings are independent
and can be processed concurrently with a single shared factorization. For $P=1$ the step
reduces to free flight; for $P=2$, $\vect M$ is $2\times2$ and
$\bar q_{1,2}=[(1+2\alpha)r_{1,2}+2\alpha r_{2,1}]/(1+4\alpha)$.

\subsection{Conditioning and stability of the solver}
\label{sec:conditioning}

The solver is numerically benign for every $P$, $\beta$ and $\dt$:
\begin{itemize}
  \item \emph{No pivoting.} $\vect M$ and $\vect T$ are symmetric and strictly
    diagonally dominant, so the Thomas recursion is backward stable without pivoting,
    with $d_k\ge1+\alpha$.\cite{golub2013,higham2002}
  \item \emph{Bounded conditioning.} The eigenvalues of $\vect M$ are
    $1+\alpha\lambda_k\in[1,1+4\alpha]$, so $\kappa_2(\vect M)\le1+(\omega_P\dt)^2$. For
    $P=128$, \SI{300}{K} and $\dt=\SI{0.1}{fs}$, $\alpha=0.063$ and $\kappa_2\le1.25$;
    even for the low-temperature dRPMD settings of \cref{sec:dtlimits},
    $\kappa_2(\vect M)\le7.3$.
  \item \emph{Well-defined correction.} By the matrix determinant lemma,
    $\det\vect M=\det\vect T\cdot\det\vect S$; since $\vect M$ and $\vect T$ are positive
    definite, $\det\vect S>0$ and $\vect S^{-1}$ always exists.
\end{itemize}

\subsection{Locality of the inverse}
\label{sec:locality}

Because $\vect M$ is circulant tridiagonal, its inverse is circulant and has the closed
form
\begin{equation}
  g_d\equiv\left(\vect M^{-1}\right)_{k,k+d}
  =\frac{r\left(r^{d}+r^{P-d}\right)}{\alpha\,(1-r^{2})(1-r^{P})},\qquad
  r=\frac{1+2\alpha-\sqrt{1+4\alpha}}{2\alpha},\qquad 0\le d\le P,
  \label{eq:banded}
\end{equation}
where $r$ is the smaller root of $\alpha r^2-(1+2\alpha)r+\alpha=0$, with $0<r<1$ and
$r\approx\alpha$ for small $\alpha$. The midpoint therefore responds to the right-hand
side with a weight that decays geometrically with bead separation: a perturbation at
one bead affects bead $k+d$ in proportion to $r^{|d|}$. The same ratio governs the
columns $\vect z_1,\vect z_2$ of \cref{eq:woodbury}, so the corner correction touches
only a few beads on either side of the seam. Truncating
$\bar q_k=\sum_{d}g_d\,r_{k+d}$, with $r_k=q_k+\tfrac12\dt\,v_k$, at $|d|\le w$,
\begin{equation}
  w=\left\lceil\frac{\ln\varepsilon}{\ln r}\right\rceil,
  \label{eq:width}
\end{equation}
reproduces $\vect M^{-1}\vect r$ to a relative accuracy of about $\varepsilon$; for
$\varepsilon=10^{-16}$ the truncation error is at the level of double-precision
round-off. The width $w$ depends only on $\alpha$, not on $P$: the Cayley step is local
in bead index, even though the normal-mode transform that defines it is global.

\section{Consequences for direct RPMD}
\label{sec:drpmd}

\subsection{dRPMD and its explicit spring integration}
\label{sec:drpmd-review}

Direct RPMD (dRPMD) is an RPMD rate method developed by Roncero and co-workers for
barrierless and complex-forming reactions at low
temperature.\cite{suleimanov2018,delmazo2019,delmazo2021} Instead of umbrella sampling and
recrossing trajectories, it simulates reactant collisions directly, as in
quasiclassical trajectory (QCT) calculations, in two stages. In the thermalization
stage the reactant separation is held fixed at a large value (\SI{120}{a_0} in
ref.~\citenum{delmazo2021}) and constrained path-integral molecular dynamics is run with
an Andersen thermostat. The relative positions and velocities of the reactants are then
reoriented and an impact parameter up to $b_\mathrm{max}$ is imposed. In the real-time
stage the constraint and thermostat are removed, the system is propagated until the
products separate or a maximum time $t_\mathrm{max}$ is reached, and the reaction
channel is assigned from the centroids. The rate coefficient is\cite{delmazo2021}
\begin{equation}
  k(T)=q_e(T)\sqrt{\frac{8k_\mathrm{B}T}{\pi\mu}}\;\pi b_{\max}^{2}(T)\,P_r(T),
  \label{eq:kdrpmd}
\end{equation}
where $q_e$ is the electronic partition-function factor, $\mu$ the reduced mass of the
reactants and $P_r$ the reaction probability.

Low-temperature dRPMD requires thousands of beads ($P=3072$ at \SI{20}{K} in
ref.~\citenum{delmazo2021}), and the code distributes beads over processes. To avoid
propagating the free ring polymer separately, which the authors expected could
aggravate spurious resonances, and to keep communication between neighbouring processes
only, dRPMD does not use the B--A--B splitting of \cref{sec:rpmd}. It instead splits the
Hamiltonian into the kinetic energy and the sum of the physical and spring potentials,
integrated together by velocity Verlet.\cite{delmazo2021} The spring forces then enter
as explicit forces, and the highest internal mode $\omega_\mathrm{max}$ and the highest
physical frequency $\Omega$ are jointly bound by the velocity-Verlet stability condition,
\begin{equation}
  \dt<\frac{2}{\sqrt{\omega_{\max}^{2}+\Omega^{2}}}\approx\frac{1}{\omega_P}=\frac{\beta\hbar}{P},\qquad
  \omega_{\max}=2\omega_P\sin\!\left(\frac{\pi\lfloor P/2\rfloor}{P}\right),
  \label{eq:vvlimit}
\end{equation}
where the approximation holds for even $P$ and $\Omega\ll\omega_P$. Because dRPMD
usually takes $P$ roughly proportional to $1/T$, $\omega_P$ is nearly the same at all
temperatures, and this bound is about \SI{0.1}{fs} throughout (\cref{sec:dtlimits});
the original authors also noted that this splitting requires a markedly smaller time
step.\cite{delmazo2021}

\subsection{Cayley integration of dRPMD}
\label{sec:drpmd-cayley}

The bead-space Cayley step meets both requirements that led dRPMD to explicit spring
integration. It operates in bead coordinates without Fourier transforms, and it treats
the spring forces by the implicit midpoint rule, which by \cref{prop:equivalence} is
stable for any $P$ and leaves only $\Omega\dt<2$. In both stages of dRPMD, velocity
Verlet is replaced by B--C--B, where B contains only the half-step impulse from the
physical potential $V$ and C is \cref{eq:system,eq:update}. During thermalization an
Andersen thermostat acts between the two B steps while the reactants are held apart, and
the real-time stage is unthermostatted B--C--B. Because C conserves $H_0$ exactly and
preserves the free ring-polymer Boltzmann distribution, thermalization still samples the
correct ring-polymer distribution, and the real-time integration remains symplectic and
time-reversible. The Hamiltonian (\cref{eq:HP}), the initial-condition protocol and the
rate expression (\cref{eq:kdrpmd}) are unchanged; only the integrator differs, and its
coefficients are computed once per temperature. The dRPMD workflow of
PyRPMD\cite{pyrpmd} is built this way, with the Cayley step as its default integrator and
velocity Verlet available as the original scheme.

\subsection{Time-step limits}
\label{sec:dtlimits}

\begin{table}[t]
  \centering
  \caption{Time-step stability limits for low-temperature dRPMD and parameters of the
  bead-space Cayley step at $\dt=\SI{0.25}{fs}$. $T$ and $P$ are those used for
  \ce{H2CO + OH} in ref.~\citenum{delmazo2021}; the velocity-Verlet (VV) limit follows
  from \cref{eq:vvlimit} with $\Omega$ corresponding to \SI{3700}{cm^{-1}}; the bound on
  $\kappa_2(\vect M)$ is $1+4\alpha$; $w$ is the truncation width of \cref{eq:width}
  for $\varepsilon=10^{-16}$; $x_1^2/12$ is the relative frequency error of the lowest
  internal mode (\cref{sec:resonance}).}
  \label{tab:dt}
  \begin{tabular}{rrcccccc}
    \toprule
    $T$/K & $P$ & $\omega_P$/fs$^{-1}$ & VV limit/fs & $\alpha$ & $\kappa_2(\vect M)\le$ & $w$ & $x_1^2/12$ \\
    \midrule
    300 &  240 &  9.43 & 0.106 & 1.39 & 6.55 & 45 & $3.2\times10^{-4}$ \\
    200 &  384 & 10.05 & 0.099 & 1.58 & 7.32 & 48 & $1.4\times10^{-4}$ \\
    100 &  768 & 10.05 & 0.099 & 1.58 & 7.32 & 48 & $3.5\times10^{-5}$ \\
     50 & 1536 & 10.05 & 0.099 & 1.58 & 7.32 & 48 & $8.8\times10^{-6}$ \\
     20 & 3072 &  8.04 & 0.124 & 1.01 & 5.04 & 39 & $1.4\times10^{-6}$ \\
    \bottomrule
  \end{tabular}
\end{table}

\Cref{tab:dt} evaluates these limits for the temperatures and bead numbers of the
low-temperature \ce{H2CO + OH} calculations of ref.~\citenum{delmazo2021}. Because $P$
scales roughly as $1/T$, $\omega_P$ varies only between 8.0 and \SI{10.1}{fs^{-1}} from
300 to \SI{20}{K}, and the velocity-Verlet limit stays at \SIrange{0.10}{0.12}{fs}. The
Cayley limit is set by the highest physical frequency alone: taking the OH stretch
(about \SI{3700}{cm^{-1}}), $\Omega\dt<2$ gives $\dt<\SI{2.9}{fs}$, 23--29 times the
velocity-Verlet limit, independent of temperature and bead number.

Both figures are stability limits; accuracy sets the production time step. The
difference lies in what imposes the limit. Under velocity Verlet, the highest internal
mode alone pins the time step near \SI{0.1}{fs}, a restriction that comes from the
internal degrees of freedom of the ring polymer rather than from the physical motion.
The Cayley step returns the limit to the physical degrees of freedom, so the time step
of dRPMD can be chosen by the same accuracy criteria as in QCT on the same potential;
\cref{sec:drpmd-numerics} quantifies the gain at \SI{20}{K}. At $\dt=\SI{0.25}{fs}$,
$\alpha=1.0$--1.6 and $\kappa_2(\vect M)\le7.3$, so the linear solve remains well
conditioned at every temperature.

\subsection{Spurious resonances are left untouched}
\label{sec:resonance}

Two kinds of resonance should be distinguished. The numerical resonance of
\cref{sec:cayley} arises when exact normal-mode rotations are combined with force
impulses near $\omega_k\dt\approx n\pi$, and the Cayley step removes it. The long-lived
trapping observed in dRPMD below \SI{100}{K} is attributed instead to spurious
resonances of the RPMD dynamics itself, in which low-lying ring-polymer internal modes
($\omega_1\approx2\pi/(\beta\hbar)$) approach the low-frequency modes of the
complex.\cite{delmazo2021,witt2009,rossi2014} This is a property of the Hamiltonian
dynamics that persists as $\dt\to0$, independent of the integrator. The Cayley step
changes the per-step rotation of an internal mode from $x_k$ to $2\arctan(x_k/2)$, a
relative frequency error of
\begin{equation}
  1-\frac{2\arctan(x_k/2)}{x_k}=\frac{x_k^2}{12}+\mathcal O(x_k^4),
  \label{eq:freqerror}
\end{equation}
which is negligible for the low-lying modes: for all settings in \cref{tab:dt} it is at
most $3.2\times10^{-4}$ for $\dt\le\SI{0.25}{fs}$ and $1.3\times10^{-3}$ for
$\dt=\SI{0.5}{fs}$. The Cayley step thus neither introduces nor masks these physical
resonances; it compresses only the frequencies of the high internal modes, whose
equilibrium distribution it still preserves exactly. Analyses of spurious resonances and
trapping in dRPMD therefore carry over unchanged.

\subsection{Bead-parallel execution}
\label{sec:parallel}

Codes that distribute the $3N$ Cartesian rings over threads, as PyRPMD does, use the
solver of \cref{sec:algorithm} unchanged. dRPMD instead assigns blocks of beads to
processes, whereas the Thomas recursion is sequential along the ring. The locality of
$\vect M^{-1}$ (\cref{sec:locality}) resolves this. Each process evaluates the truncated
convolution $\bar q_k=\sum_{|d|\le w}g_d\,r_{k+d}$ for its own beads and for one boundary
bead on each side, which requires the positions and velocities of $w+1$ boundary beads
from each neighbouring process; \cref{eq:update} then needs no further communication.
When $2w+1\ge P$ the full solver of \cref{sec:algorithm} is used instead. Provided each
process holds at least $w+1$ beads, communication remains between neighbouring processes
only, as in the original dRPMD scheme; only the halo widens from one bead to $w+1$. For
the settings of \cref{tab:dt}, $w=39$--48 is far below $P$, so 40--49 beads per process
suffice. The coefficients $g_d$ are mass-independent and computed once per $P$, $\beta$
and $\dt$. The truncated convolution costs about $2(2w+1)$ floating-point operations per
bead and degree of freedom, more than the Thomas recursion, but has no sequential
dependence and vectorizes readily.

\section{Numerical verification}
\label{sec:numerics}

\subsection{Implementation and test coverage}
\label{sec:implementation}

The bead-space Cayley step is the default integrator (\code{cayley-bcb}) of
PyRPMD,\cite{pyrpmd} implemented in C++17 with OpenMP (\code{native/cayley.cpp}). Each
propagator object holds one immutable factorization (step 1 of \cref{sec:algorithm}),
the $3N$ Cartesian rings are distributed over threads, and $P=1$ and $P=2$ use the
special cases above. The same compiled module provides the exact normal-mode B--A--B
step (\code{normal-mode-bab}, with a radix-2 FFT for power-of-two $P\ge32$) and
full-Hamiltonian velocity Verlet (\code{velocity-verlet}) behind the same interface, and
an independent NumPy reference evaluates the Cayley map in Fourier space. All three
schemes drive the same rate, dRPMD and path-integral workflows, so the comparisons below
differ only in the integrator. The integrator test suite checks every scheme against a
dense matrix exponential of the harmonic ring-polymer Hamiltonian and explicit
graph-Laplacian forces; it covers $P=1$, odd and even rings, second-order convergence,
time reversibility, conservation of the free ring-polymer energy at time steps beyond
the velocity-Verlet limit, and enforcement of that limit (the 277 tests of the five modules that exercise these integrators pass at commit
\code{8f32b37}). The scripts that produce
the numbers below are provided with this article.

\subsection{Agreement with normal-mode Cayley to machine precision}

For six atoms at \SI{300}{K} and $\dt=\SI{0.1}{fs}$, we propagated six random states
by one step with the native bead-space solver, with the NumPy normal-mode Cayley step,
and with an independent dense solve of \cref{eq:system,eq:update}. Over $P=4$--2048, the
maximum relative deviation from normal-mode Cayley is $1.5\times10^{-15}$ and from the
dense solve $2.3\times10^{-15}$, as \cref{prop:equivalence} requires.

Errors do not accumulate over long runs. Over $2\times10^5$ consecutive steps
(\SI{20}{ps}) from three random states, the free ring-polymer energy stays within
$3\times10^{-11}$ of its initial value for $P=32$ and within $6\times10^{-11}$ for
$P=128$, and the centroid deviates from exact free flight by at most
$4\times10^{-11}$ in relative terms. These deviations are accumulated round-off: the
bead-space and normal-mode propagators generate the same trajectories.

\subsection{Unconditional stability}

\begin{figure}[t]
  \centering
  \includegraphics[width=0.62\textwidth]{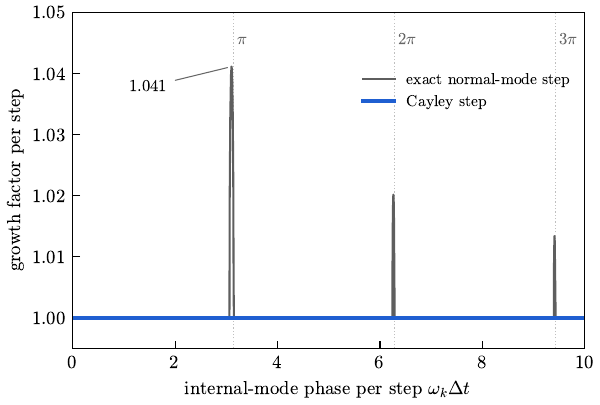}
  \caption{Growth factor per step (spectral radius of the one-step B--A--B map) of an
  internal mode under a harmonic external force with $\Omega\dt=0.5$, from the
  half-traces of \cref{sec:cayley}. The exact normal-mode step is unstable just below
  every $n\pi$, with a peak of 1.041 per step; the Cayley step has unit growth factor at
  every phase.}
  \label{fig:stability}
\end{figure}

\Cref{fig:stability} evaluates the one-step map for a harmonic external force. The
exact normal-mode step develops growth factors above unity as the internal-mode phase
approaches $\pi$, $2\pi$ and $3\pi$, peaking at 1.041 per step, an amplification of
about $3\times10^{17}$ after 1000 steps. The Cayley step, and hence the bead-space step,
has a growth factor of exactly one at every phase. Provided $\Omega\dt<2$, it is stable
for any number of beads, and the condition no longer tightens with $P$.

\subsection{Cost}
\label{sec:cost}

\begin{figure}[t]
  \centering
  \includegraphics[width=\textwidth]{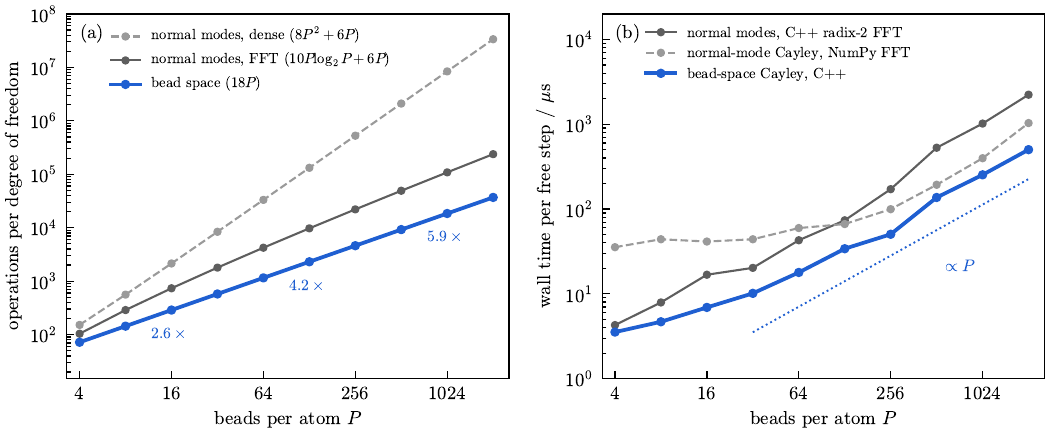}
  \caption{Cost of one free ring-polymer step.
  (a)~Floating-point operations per degree of freedom: bead-space Cayley ($18P$,
  \cref{sec:algorithm}), normal-mode step with four real FFTs ($10P\log_2P+6P$), and with
  a dense transform ($8P^2+6P$); numbers give the ratio of the FFT count to the
  bead-space count.
  (b)~Wall time in PyRPMD for six atoms at \SI{300}{K} and $\dt=\SI{0.1}{fs}$ (single
  \SI{2.1}{GHz} Xeon core, median of seven repeats, including the Python call):
  bead-space Cayley and the normal-mode step of the same C++ module, and the NumPy
  normal-mode Cayley step. The dotted line is proportional to $P$.}
  \label{fig:cost}
\end{figure}

The bead-space step needs fewer operations than the FFT-based normal-mode step for every
$P\ge4$, and the margin widens as $\log_2P$ (\cref{fig:cost}a): the ratio is 4.2 at
$P=128$ and 6.4 at $P=2048$. Against a dense transform the ratio grows linearly in $P$.
The measured cost follows the same pattern (\cref{fig:cost}b). In PyRPMD the bead-space
step grows linearly with $P$, from \SI{34}{\micro s} at $P=128$ to \SI{501}{\micro s} at
$P=2048$ for six atoms. It is 2.0--4.4 times faster than the normal-mode step compiled
in the same module for $P\ge32$, and 1.4--2.1 times faster than the NumPy normal-mode
Cayley step for $P\ge128$, where the per-call overhead of the NumPy path no longer
dominates. Because the comparison in the same module shares the language, compiler,
memory layout and threading, the measured gap reflects the transform stage that the
bead-space formulation removes. The resulting speed-up of a complete RPMD simulation
depends on the share of the free ring-polymer step in the total cost, which is largest
for inexpensive potentials and large bead numbers.

\subsection{Time step of dRPMD at 20 K}
\label{sec:drpmd-numerics}

\begin{figure}[t]
  \centering
  \includegraphics[width=0.62\textwidth]{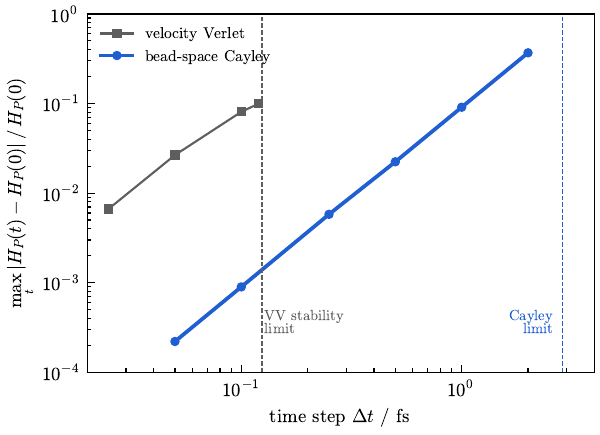}
  \caption{Largest relative deviation of $H_P$ from its initial value over \SI{1}{ps}
  for $P=3072$ at \SI{20}{K}, for an OH-like oscillator ($\Omega=\SI{3700}{cm^{-1}}$,
  OH reduced mass) propagated by PyRPMD with velocity Verlet and with the bead-space
  Cayley step from the same thermal initial state. Dashed lines mark the stability
  limits of \cref{tab:dt}: \SI{0.124}{fs} for velocity Verlet and \SI{2.87}{fs} for
  Cayley.}
  \label{fig:drpmd}
\end{figure}

To test what the Cayley step gains in practice, we propagated the \SI{20}{K}, $P=3072$
setting of \cref{tab:dt} with both integrators. The physical potential is an isotropic
harmonic well with the OH-stretch frequency and the OH reduced mass, which represents the
stiffest physical motion assumed in \cref{tab:dt}. The ring starts from the exact thermal
distribution of the free ring polymer, the centroid from the classical distribution in
the well, and the momenta from the Maxwell distribution, and every run starts from the
same state. \Cref{fig:drpmd} shows the largest relative deviation of $H_P$ over
\SI{1}{ps}, the quantity monitored in dRPMD production runs.

The Cayley step conserves $H_P$ to the same fidelity as velocity Verlet with a time step
about ten times larger: $5.8\times10^{-3}$ at \SI{0.25}{fs} against $6.7\times10^{-3}$ for
velocity Verlet at \SI{0.025}{fs}, and $9.1\times10^{-2}$ at \SI{1}{fs} against
$8.1\times10^{-2}$ at \SI{0.1}{fs}. At equal time step the deviation is 90--120 times
smaller, because the energy error of velocity Verlet comes from the stiff internal modes
of the ring, whereas that of the Cayley step comes from the physical force alone. PyRPMD
rejects velocity Verlet above \SI{0.124}{fs}, the limit of \cref{eq:vvlimit}, while
the Cayley step remains stable up to \SI{2}{fs}, the largest step tested. In this
setting, a given standard of energy conservation is therefore reached with about one
tenth of the force evaluations.

\section{Conclusions}
\label{sec:conclusions}

In bead coordinates, the Cayley step of RPMD is a single nearest-neighbour linear system
for the midpoint of the step, with a matrix that depends only on $P$, $\beta$ and $\dt$.
It is exactly the normal-mode Cayley map, so it conserves the free ring-polymer energy,
preserves its Boltzmann distribution, and is stable for any number of beads whenever
$\Omega\dt<2$. It is solved by a pivot-free $\mathcal O(P)$ recursion with a condition
number bounded by $1+(\omega_P\dt)^2$, and its inverse decays geometrically with bead
separation, so the step is local in bead index. As the default integrator of PyRPMD, it
reproduces normal-mode Cayley to machine precision for up to 2048 beads, and its cost
grows linearly with $P$ and stays below that of the FFT-based normal-mode step.

These properties make the Cayley step available wherever normal-mode transforms are
inconvenient. For dRPMD, which has integrated the springs explicitly with a time step
bounded by $\beta\hbar/P$, they raise the stability limit 23--29-fold in published
low-temperature settings, to a bound set by the physical frequencies alone; at
\SI{20}{K} with 3072 beads, the same energy conservation is reached with a time step
about ten times larger. The low-lying internal modes, and hence the analysis of spurious
resonances and trapping, are left essentially unchanged, and communication remains
between neighbouring processes. More generally, any stable integration scheme whose
per-mode coefficients are rational functions of the squared frequency can be recast in
the same way as a banded system in bead coordinates, which opens the same route for other
strongly stable ring-polymer integrators.

\section*{Code availability}
The bead-space Cayley propagator is implemented in PyRPMD (commit \code{8f32b37}),
available at \url{https://github.com/chen-yu-hao/PyRPMD}. The scripts and
raw outputs behind \cref{fig:cost,fig:drpmd} and the numbers in \cref{sec:numerics} are
included with this article.

\section*{Acknowledgments}
X.H.\ was supported by the National Natural Science Foundation of China (Grant
Nos.\ 22633005, 92477103 and 22273023), Shanghai Municipal Science and Technology
Commission with Grant No.\ 25511102400, Shanghai Municipal Commission of Economy and
Informatization with Grant No.\ 2026-GZL-RGZN-01015, Advanced Materials-National Science
and Technology Major Project (Grant No.\ 2026ZD0623702), the Shanghai Frontiers Science
Center of Molecule Intelligent Syntheses, and the Fundamental Research Funds for the
Central Universities. We also acknowledge the Supercomputer Center of East China Normal
University (ECNU Multifunctional Platform for Innovation 001) for providing computer
resources.

\bibliographystyle{achemso}
\bibliography{refs}

@article{craig2004,
  author  = {Craig, Ian R. and Manolopoulos, David E.},
  title   = {Quantum statistics and classical mechanics: Real time correlation functions from ring polymer molecular dynamics},
  journal = {J. Chem. Phys.},
  year    = {2004},
  volume  = {121},
  pages   = {3368}
}

@article{craig2005a,
  author  = {Craig, Ian R. and Manolopoulos, David E.},
  title   = {Chemical reaction rates from ring polymer molecular dynamics},
  journal = {J. Chem. Phys.},
  year    = {2005},
  volume  = {122},
  pages   = {084106}
}

@article{craig2005b,
  author  = {Craig, Ian R. and Manolopoulos, David E.},
  title   = {A refined ring polymer molecular dynamics theory of chemical reaction rates},
  journal = {J. Chem. Phys.},
  year    = {2005},
  volume  = {123},
  pages   = {034102}
}

@article{habershon2013,
  author  = {Habershon, Scott and Manolopoulos, David E. and Markland, Thomas E. and Miller, III, Thomas F.},
  title   = {Ring-polymer molecular dynamics: Quantum effects in chemical dynamics from classical trajectories in an extended phase space},
  journal = {Annu. Rev. Phys. Chem.},
  year    = {2013},
  volume  = {64},
  pages   = {387}
}

@article{suleimanov2016,
  author  = {Suleimanov, Yury V. and Aoiz, F. Javier and Guo, Hua},
  title   = {Chemical reaction rates from ring polymer molecular dynamics: Theory and practical applications},
  journal = {J. Phys. Chem. A},
  year    = {2016},
  volume  = {120},
  pages   = {8488},
  doi     = {10.1021/acs.jpca.6b07140}
}

@article{ceriotti2010pile,
  author  = {Ceriotti, Michele and Parrinello, Michele and Markland, Thomas E. and Manolopoulos, David E.},
  title   = {Efficient stochastic thermostatting of path integral molecular dynamics},
  journal = {J. Chem. Phys.},
  year    = {2010},
  volume  = {133},
  pages   = {124104}
}

@article{korol2019,
  author  = {Korol, Roman and Bou-Rabee, Nawaf and Miller, III, Thomas F.},
  title   = {Cayley modification for strongly stable path-integral and ring-polymer molecular dynamics},
  journal = {J. Chem. Phys.},
  year    = {2019},
  volume  = {151},
  pages   = {124103}
}

@article{korol2020,
  author  = {Korol, Roman and Rosa-Ra{\'\i}ces, Jorge L. and Bou-Rabee, Nawaf and Miller, III, Thomas F.},
  title   = {Dimension-free path-integral molecular dynamics without preconditioning},
  journal = {J. Chem. Phys.},
  year    = {2020},
  volume  = {152},
  pages   = {104102}
}

@article{rosaraices2021,
  author  = {Rosa-Ra{\'\i}ces, Jorge L. and Sun, Jiace and Bou-Rabee, Nawaf and Miller, III, Thomas F.},
  title   = {A generalized class of strongly stable and dimension-free {T-RPMD} integrators},
  journal = {J. Chem. Phys.},
  year    = {2021},
  volume  = {154},
  pages   = {024106},
  doi     = {10.1063/5.0036954}
}

@article{gui2022,
  author  = {Gui, Xiongfei and Fan, Wenbin and Sun, Jiace and Li, Yongle},
  title   = {New stable and fast ring-polymer molecular dynamics for calculating bimolecular rate coefficients with an example of {OH} + {CH$_4$}},
  journal = {J. Chem. Theory Comput.},
  year    = {2022},
  volume  = {18},
  pages   = {5203--5212},
  doi     = {10.1021/acs.jctc.2c00522}
}

@article{collepardo2009,
  author  = {Collepardo-Guevara, Rosana and Suleimanov, Yury V. and Manolopoulos, David E.},
  title   = {Bimolecular reaction rates from ring polymer molecular dynamics},
  journal = {J. Chem. Phys.},
  year    = {2009},
  volume  = {130},
  pages   = {174713},
  note    = {Erratum: \textit{J. Chem. Phys.} \textbf{2010}, \textit{133}, 049902}
}

@article{suleimanov2013,
  author  = {Suleimanov, Yury V. and Allen, Joshua W. and Green, William H.},
  title   = {{RPMDrate}: Bimolecular chemical reaction rates from ring polymer molecular dynamics},
  journal = {Comput. Phys. Commun.},
  year    = {2013},
  volume  = {184},
  pages   = {833--840},
  doi     = {10.1016/j.cpc.2012.10.017}
}

@article{andersen1980,
  author  = {Andersen, Hans C.},
  title   = {Molecular dynamics simulations at constant pressure and/or temperature},
  journal = {J. Chem. Phys.},
  year    = {1980},
  volume  = {72},
  pages   = {2384}
}

@article{ceriotti2010gle,
  author  = {Ceriotti, Michele and Bussi, Giovanni and Parrinello, Michele},
  title   = {Colored-noise thermostats {\`a} la carte},
  journal = {J. Chem. Theory Comput.},
  year    = {2010},
  volume  = {6},
  pages   = {1170}
}

@book{hairer2006,
  author    = {Hairer, Ernst and Lubich, Christian and Wanner, Gerhard},
  title     = {Geometric Numerical Integration: Structure-Preserving Algorithms for Ordinary Differential Equations},
  edition   = {2nd},
  publisher = {Springer},
  address   = {Berlin},
  year      = {2006}
}

@article{temperton1975,
  author  = {Temperton, Clive},
  title   = {Algorithms for the solution of cyclic tridiagonal systems},
  journal = {J. Comput. Phys.},
  year    = {1975},
  volume  = {19},
  pages   = {317--323}
}

@article{hager1989,
  author  = {Hager, William W.},
  title   = {Updating the inverse of a matrix},
  journal = {SIAM Rev.},
  year    = {1989},
  volume  = {31},
  pages   = {221--239}
}

@book{golub2013,
  author    = {Golub, Gene H. and Van Loan, Charles F.},
  title     = {Matrix Computations},
  edition   = {4th},
  publisher = {Johns Hopkins University Press},
  address   = {Baltimore},
  year      = {2013}
}

@book{higham2002,
  author    = {Higham, Nicholas J.},
  title     = {Accuracy and Stability of Numerical Algorithms},
  edition   = {2nd},
  publisher = {SIAM},
  address   = {Philadelphia},
  year      = {2002}
}

@article{suleimanov2018,
  author  = {Suleimanov, Yury V. and Aguado, Alfredo and G{\'o}mez-Carrasco, Susana and Roncero, Octavio},
  title   = {A ring polymer molecular dynamics approach to study the transition between statistical and direct mechanisms in the {H$_2$} + {H$_3^+$} $\to$ {H$_3^+$} + {H$_2$} reaction},
  journal = {J. Phys. Chem. Lett.},
  year    = {2018},
  volume  = {9},
  pages   = {2133}
}

@article{delmazo2019,
  author  = {del Mazo-Sevillano, Pablo and Aguado, Alfredo and Jim{\'e}nez, Elena and Suleimanov, Yury V. and Roncero, Octavio},
  title   = {Quantum roaming in the complex-forming mechanism of the reactions of {OH} with formaldehyde and methanol at low temperature and zero pressure: A ring polymer molecular dynamics approach},
  journal = {J. Phys. Chem. Lett.},
  year    = {2019},
  volume  = {10},
  pages   = {1900}
}

@article{delmazo2021,
  author  = {del Mazo-Sevillano, Pablo and Aguado, Alfredo and Roncero, Octavio},
  title   = {Neural network potential energy surface for the low temperature ring polymer molecular dynamics of the {H$_2$CO} + {OH} reaction},
  journal = {J. Chem. Phys.},
  year    = {2021},
  volume  = {154},
  pages   = {094305},
  doi     = {10.1063/5.0044009}
}

@article{witt2009,
  author  = {Witt, Alexander and Ivanov, Sergei D. and Shiga, Motoyuki and Forbert, Harald and Marx, Dominik},
  title   = {On the applicability of centroid and ring polymer path integral molecular dynamics for vibrational spectroscopy},
  journal = {J. Chem. Phys.},
  year    = {2009},
  volume  = {130},
  pages   = {194510}
}

@article{rossi2014,
  author  = {Rossi, Mariana and Ceriotti, Michele and Manolopoulos, David E.},
  title   = {How to remove the spurious resonances from ring polymer molecular dynamics},
  journal = {J. Chem. Phys.},
  year    = {2014},
  volume  = {140},
  pages   = {234116}
}

@misc{pyrpmd,
  author       = {Chen, Yuhao},
  title        = {{PyRPMD}: Ring-polymer molecular dynamics, version 0.2.0},
  year         = {2026},
  howpublished = {\url{https://github.com/chen-yu-hao/PyRPMD}},
  note         = {Commit 8f32b37}
}

\end{document}